\documentclass{copernicus} 

\nolinenumbers
\begin{document}
\title{Uncertainties in the heliosheath ion temperatures}

\Author[1,2]{Klaus}{Scherer}
\Author[3]{Hans J\"org}{Fahr} 
\Author[1,2]{Horst}{Fichtner}
\Author[1]{Adama}{Sylla}
\Author[4]{John D.}{Richardson}
\Author[1]{Marian}{Lazar}
\affil[1]{Institut f\"ur Theoretische Physik IV: Weltraum- und
  Astrophysik, Ruhr-Universit\"at Bochum, Germany}
\affil[2]{Research Department, Plasmas with Complex Interactions,
  Ruhr-Universit\"at Bochum, Germany}
\affil[3]{Argelander Institut f\"{u}r Astronomie, Universit\"{a}t Bonn, Germany}
\affil[4]{MIT}
\runningtitle{Heliosheath temperatures}
\runningauthor{Scherer et al.}
\correspondence{Klaus Scherer (kls@tp4.rub.de)}
\received{}
\revised{}
\accepted{}
\published{}
\firstpage{1}
\renewcommand{\bf}{}
\maketitle

\begin{abstract}
  The Voyager plasma observations show that the physics of the
  heliosheath is rather complex, and especially that the temperature
  derived from observation differs from expectations.  To explain this
  fact the temperature in the heliosheath should be based on $\kappa$
  distributions instead of Maxwellians {\bf because the former} allows
  for much higher temperature. Here we show an easy way to calculate
  the $\kappa$
  temperatures when those estimated from the data are given as
  Maxwellian temperatures.  We use the moments of the Maxwellian and
  $\kappa$
  distributions to estimate the $\kappa$
  temperature. Moreover, we show that the pressure (temperature) given
  by a truncated $\kappa$
  distribution is similar to that given by a Maxwellian and only
  starts to increase for higher truncation velocities.  We deduce a
  simple formula to convert the Maxwellian to $\kappa$
  pressure or temperature. We apply this result to the Voyager-2
  observations in the heliosheath.
\end{abstract}
%
\introduction
Knowledge about the temperature of an astrophysical plasma is of
significance for the correct hydrodynamical treatment of various
plasma processes like, e.g., heat conduction, wave propagation,
compression or charge exchange. The temperature of a space plasma
is, of course, not directly measurable but must be derived
indirectly. If the velocity distribution function of a plasma
constituent is known, the temperature can be computed as a second
velocity moment. If not, assumptions have to be made about this
distribution function. Consequently, an uncertainty regarding the
latter will translate into an uncertainty regarding the
temperature. For simplicity, unmeasured velocity distributions are
often assumed to be Maxwellians. One must be aware that the derived
`Maxwellian temperature' is an assumption and might be rather
different from the actual thermodynamically relevant temperature of
the considered plasma
\citep[e.g.,][]{Fahr-Siewert-2013,Nicolaou-Livadiotis-2016}.

One such example is the plasma in the inner heliosheath, i.e.\ the
region of the heliosphere between the solar wind termination shock and
the heliopause.  The temperatures for the inner heliosheath can, in
difference to those determined for the upstream solar wind
\citep{Bridge-etal-1977}, only be derived from Voyager~2 measurements
under the Maxwellian assumption
\citep[e.g.,][]{Richardson-Wang-2012}. From the modeling of fluxes of
energetic neutral atoms (ENAs) it is known, however, that the
distribution function of protons in this region is probably not a
Maxwellian but a so-called $\kappa$-distribution with the parameter
$\kappa < 2$ \citep[e.g.][]{Heerikhuisen-etal-2008}. Consequently,
the temperature of the proton population should be computed from these
non-Maxwellian distributions. In the present paper we quantify these
alternative $\kappa$-temperatures and discuss their significant
difference to the Maxwellian ones. Before both temperatures are
discussed in detail, we briefly review the $\kappa$-distributions and
the physical reason for their importance downstream of the termination
shock.

\section{$\ensuremath{\kappa}$-distributions}
The $\kappa$-distributions
are an often used tool for the quantitative treatment of
non-Maxwellian plasmas as is described in the reviews by
\citet{Pierrard-Lazar-2010} and \citet{Livadiotis-McComas-2013}.
{\bf In the following, we assume that the actual distribution function
  is isotropic in the bulk velocity frame, because the protons undergo
  efficient pitch angle scattering at Alfv{\'e}n waves (MHD wave
  turbulences) leading to this form of isotropy. Furthermore, without
  the capabilities to measure 2D or 3D distributions and implicitly
  their anisotropy, the observations always integrated over all
  pitch-angles, and thus isotropic.  Thus, the definition of
  the isotropic $\kappa$-distribution reads:}
\begin{eqnarray}
  \label{eq:3}
  f_{\kappa}(\vec{v}) &=& n(\vec{r}) \dfrac{\Gamma(\kappa+1)}{
  \left(\sqrt{\pi}\sqrt{\kappa}\Theta\right)^{3}\Gamma\left(\kappa-\dfrac{1}{2}\right)}
     \left(1+\dfrac{v^{2}}{\kappa\Theta^{2}}\right)^{-(\kappa+1)}
\end{eqnarray}
where $\Theta$
is a fitting parameter, which normalises the speed, often it is
identified with a thermal speed.  In the limit $\kappa\to\infty$
these distributions converge to the Maxwellian distribution:
\begin{eqnarray}
  \label{eq:1}
  f_{m}(\vec{v}) &=& n_{0}(\vec{r}) \left(\frac{m}{2 \pi k T}\right)^{3/2}
\exp\left(-\frac{m\vec{v}^{2}}{2 k T}\right) \qquad \mathrm{or}\\\nonumber
 &=& n_{0}(\vec{r})\left(\pi v_{p}^{2}\right)^{-3/2} \exp \left(-\frac{v^{2}}{v_{p}^{2}}\right)
\end{eqnarray}
where the fitting parameter is usually the `thermal speed'
$v_{p} \equiv \sqrt{2 kT /m}$,
with the temperature $T$,
the proton mass $m$ and the Boltzmann constant $k$.

Both distribution functions are normalized to the number density
$n(\vec{r})$ so that
\begin{eqnarray}
  \label{eq:2}
  \int f_{m,\kappa}d^{3}v=4\pi\int\limits_{0}^{\infty} v^{2}f_{m,\kappa}dv=n(\vec{r})
\end{eqnarray}
Furthermore, for the following, we normalize all speeds ($v, \Theta$) to the
solar wind speed $u_{sw}=436$\,km/s which corresponds to a proton energy of 1\,keV.

{\bf In order to fit the
thermal core of the Maxwellian,  it is required that the speed $\Theta$
is equal to the thermal speed $\Theta=v_p$, as 
shown in appendix~A.} The latter relation 
corresponds to the preferable of two alternatives to interpret
$\kappa$-temperatures, as is discussed in detail in \citet{Lazar-etal-2015}
and \citet{Lazar-etal-2016}. 

\subsection{Pick-up proton induced $\ensuremath{\kappa}$-distributions}
Due to the different velocity-space processes acting upon solar wind
protons when moving out from 1~AU to great radial solar distances, the
resulting distribution function is far from being an equilibrium
distribution in the form of a shifted Maxwellian. This is
theoretically evident mainly from the fact that the solar wind proton
plasma is permanently loaded with newly injected pick-up
protons. This, in connection with wave-driven diffusion processes,
keeps the resulting ion distribution function off a Maxwellian
equilibrium shape. The transport of such pick-up ions was a subject of
many investigations in the past, see \citet{Fahr-1973},
\citet{Holzer-Leer-1973}, and \citet{Vasyliunas-Siscoe-1976} with an
ever increasing sophistication in treating the exact form of this
pick-up ion incorporation process
\citep{Isenberg-1995, Fichtner-etal-1996, Schwadron-etal-1996,
  Chalov-Fahr-1998, Pogorelov-etal-2016}.
Also from in-situ observations it had been clearly recognized that
pick-up protons show a typical core-distribution below the velocity
injection border and an extended power-law tail above
\citep{Gloeckler-2003, Fisk-etal-2010, Hill-etal-2009,
  McComas-etal-2015b}.
This then justified the theoretical endeavor of \citet{Fahr-etal-2014}
to treat the evolving joint solar wind ion distribution as a $\kappa$-distribution with
a $\kappa$-parameter evolving with radial distance. These authors could show with the
help of an adequate pressure-moment transport equation that the
resulting ion kappa-distribution evolves into a highly non-thermal
distribution with $\kappa\leq 2.0$ by the time the solar wind plasma
arrives at the termination shock.

From this finding it became evident that the passage of such a
non-equilibrium proton distribution over the solar wind termination shock
would generate a  distribution downstream of the shock with
strong non-equilibrium signatures. There are good reasons given by
the Liouville theorem, see \citet{Siewert-Fahr-2007} that this
downstream distribution function will also be as a kappa function obeying
the relation
\begin{eqnarray}
  \label{eq:h1}
  f_{2}(v)=s\cdot f_{1}(\frac{v}{\sqrt{s}}) 
\end{eqnarray}
where the indices "1" and "2" indicate upstream and downstream quantities,
respectively, and where $s$ denotes the shock compression ratio. This
then shows that downstream of the shock the following kappa function has\ to
be expected
\begin{eqnarray}
  \label{eq:h2}
f_{2}(v) = A s  \left(1 + \frac{v^{2}}{\kappa
_{1}s \Theta _{1}^{2}}\right)^{-(\kappa _{1}+1)} 
\end{eqnarray}
This expresses the fact that downstream one must expect a kappa distribution with
$\kappa$ identical to its upstream value $\kappa_{1}$, but with an
increased core width $\Theta _{2}=s\Theta _{1}$.

The above finding raises the question how in-situ low-energy ion
measurements, e.g. by Voyager-2, should be interpreted in terms of
velocity-moments of the distribution function. Solar wind ion data
downstream of the shock are, in general, not displayed as spectral
flux ion data, but as plasma parameters ($n$, flow speed, and $T$)
derived from a fit of the measured fluxes to a convected isotropic
Maxwellian distribution, see e.g.\ \citet{Richardson-Wang-2012}. The
question arises on the basis of whether the underlying distribution
function is a Maxwellian and if not how the actual plasma parameters
differ from those derived using a Maxwellian fit.  A recently
published paper by \citet{Nicolaou-Livadiotis-2016} demonstrated, how
different values of the velocity moments can be, if based on
Maxwellian or on $\kappa$-distributions. In the following part of the paper we shall
thus focus on the specific question, how Voyager-2 heliosheath plasma
parameters displayed in the literature change values when interpreted
on the basis of non-thermal ion kappa distributions.

\section{Maxwellian temperatures along the Voyager-2 trajectory}

\begin{figure}[t!]
  \includegraphics[width=0.99\columnwidth]{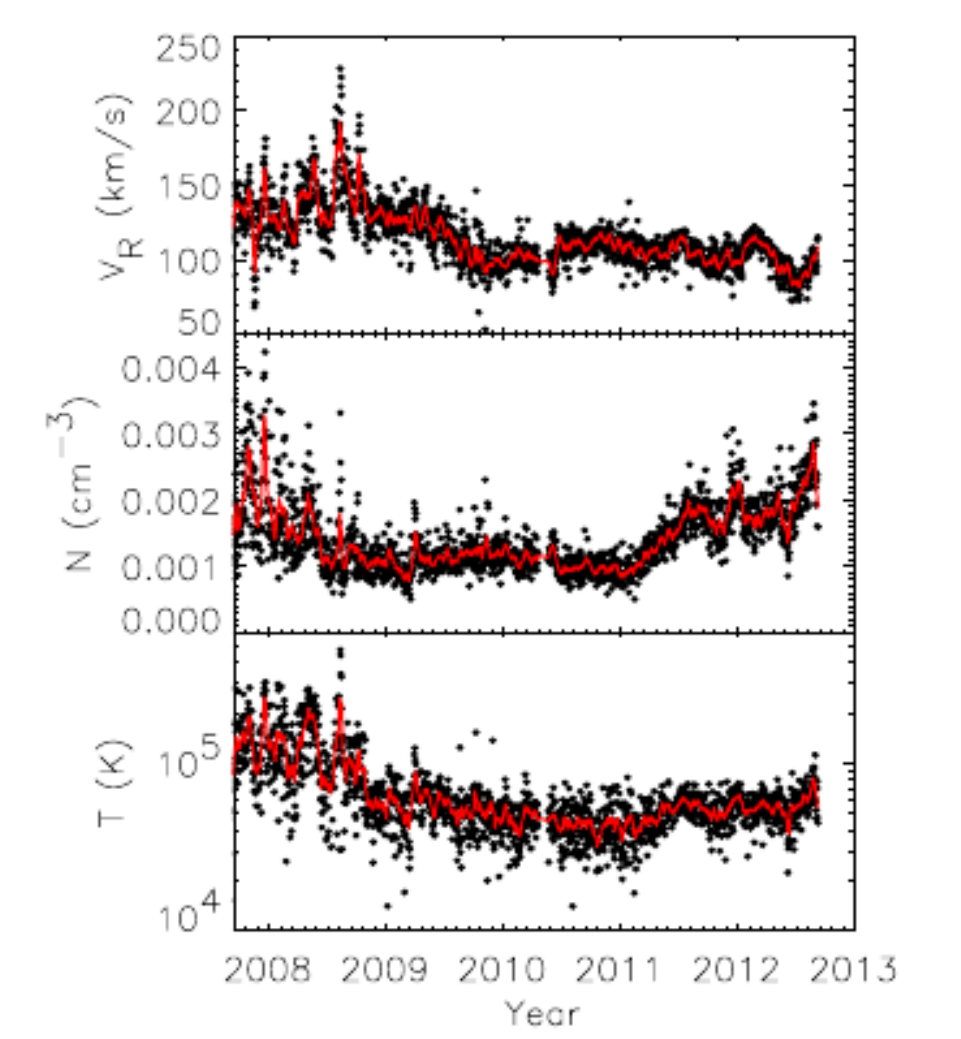}  
  \caption{Daily (points) and 11 day running averages (lines) of the
    radial speed, density, and temperature observed at V2. Taken from
    \citet{Richardson-Wang-2012}. }
  \label{fig:4}
\end{figure}

\citet{Richardson-Wang-2012} have shown that during the years 2008 to
2011 (denoted by $t_{8}$ and $t_{11}$ in the following text) the
heliosheath proton temperature measured along the Voyager-2 trajectory has
fallen from $T(t_{8})=140.000$\,K to $T(t_{11})=40.000$\,K (see
Fig.~\ref{fig:4}). This temperature decrease occurred while the
measured proton density remained fairly constant at an average value
of $\langle n_{p}\rangle \simeq 0.001cm^{-3}$. The temperature
decrease during the Voyager-2 time period of $t_{11}-t_{8}=3$\,yrs thus
indicates a non-adiabatic origin.  We begin by investigating this
temperature decrease in a first-order view studying the temperature in
a plasma volume comoving with the plasma bulk velocity $\vec{u}_{sw}$
being subject to ongoing charge exchange reactions with cold LISM\
H-atoms.

In this volume a proper time $\tau $ is counted (i.e. the time of the
comoving clock), and we want to describe the temperature change within
the box as function of $\tau $. For an adequate estimate we use a
simple thermodynamic transport equation describing the proton
temperature  as the result of
charge-exchange-related\ energy exchanges between the plasma and the
neutral gas in the following form:
\begin{equation}\label{hans1}
\frac{d}{d\tau }(n_{p}k T)=-n_{p}n_{H}\sigma _{ex}v_{rel}\cdot (k T-k T_{0})
\end{equation}
with $n_{p}\simeq const$ (see the assumption of incompressible
heliosheath plasma flow as made in \citet{Fahr-etal-2016}),
$n_{H}\simeq const$, $k T\gg k T_{0}$, and $T(\tau )\sim O(10^{5}K)$\
(see Richardson and Wang, 2012) denoting the actual, local ion
temperature, $T_{0}$ denoting the thermalized heliosheath pick-up ion
temperature downstream of the shock with
$T_{0}\simeq (3/2K)mU_{0}^{2}= O(10^{3}K)$ denoting the temperature of
the newly injected ions, originating from the cold LISM H-atoms (where
$U_{0}$ is the bulk speed of the proton fluid in the heliosheath).
This equation evidently can be simplified to
\begin{equation}
\frac{1}{T}\frac{d}{d\tau }(T)=-n_{H}\sigma _{ex}v_{rel}
\end{equation}
and with $v_{rel}\simeq \sqrt{8\pi  k T/m}$  leads to
\begin{equation}
\frac{1}{T^{3/2}}\frac{d}{d\tau }(T)=-n_{H}\sigma _{ex}\sqrt{8\pi k/m}
\end{equation}
yielding with $\sigma _{ex}=const$
\begin{equation}
  \frac{1}{\sqrt{T}}-\frac{1}{\sqrt{T_{8}}}=n_{H}\sigma _{ex}\sqrt{\frac{2\pi k%
    }{m}}(\tau -\tau _{8})
\end{equation}
where $\tau _{8}=t_{8}=$Jan.\,2008, the date when Voyager-2 started moving
within the heliosheath.  Adopting a time period
$(\tau _{11}-\tau _{8})\simeq (t_{11}-t_{8})=9.6\cdot 10^{7}s$ we find
\begin{eqnarray}
\sqrt{T_{11}/T_{8}}&=&\frac{1}{1+n_{H}\sigma _{ex}\sqrt{\frac{2\pi k T_{8}}{m}}
(t_{11}-t_{8})}=\frac{1}{1.19}
\end{eqnarray}
assuming $\sqrt{\frac{2\pi k T_{8}}{m}}=100$\,km/s and $n_{H}\sigma
\approx 10^{-16}cm^{-1}$.  This yields
$T_{11}=T_{8}/1.119^{2}\approx 988630$\,K.  This estimate has to be
compared with temperatures of about $40000K$ were measured by Voyager-2 in
2011 and indicates that an essential part of the ion cooling may be
ascribed to ongoing charge exchange reactions removing high energy
ions and replacing them by low energetic ones. 

\begin{figure}[t!]
  \includegraphics[width=0.9\columnwidth]{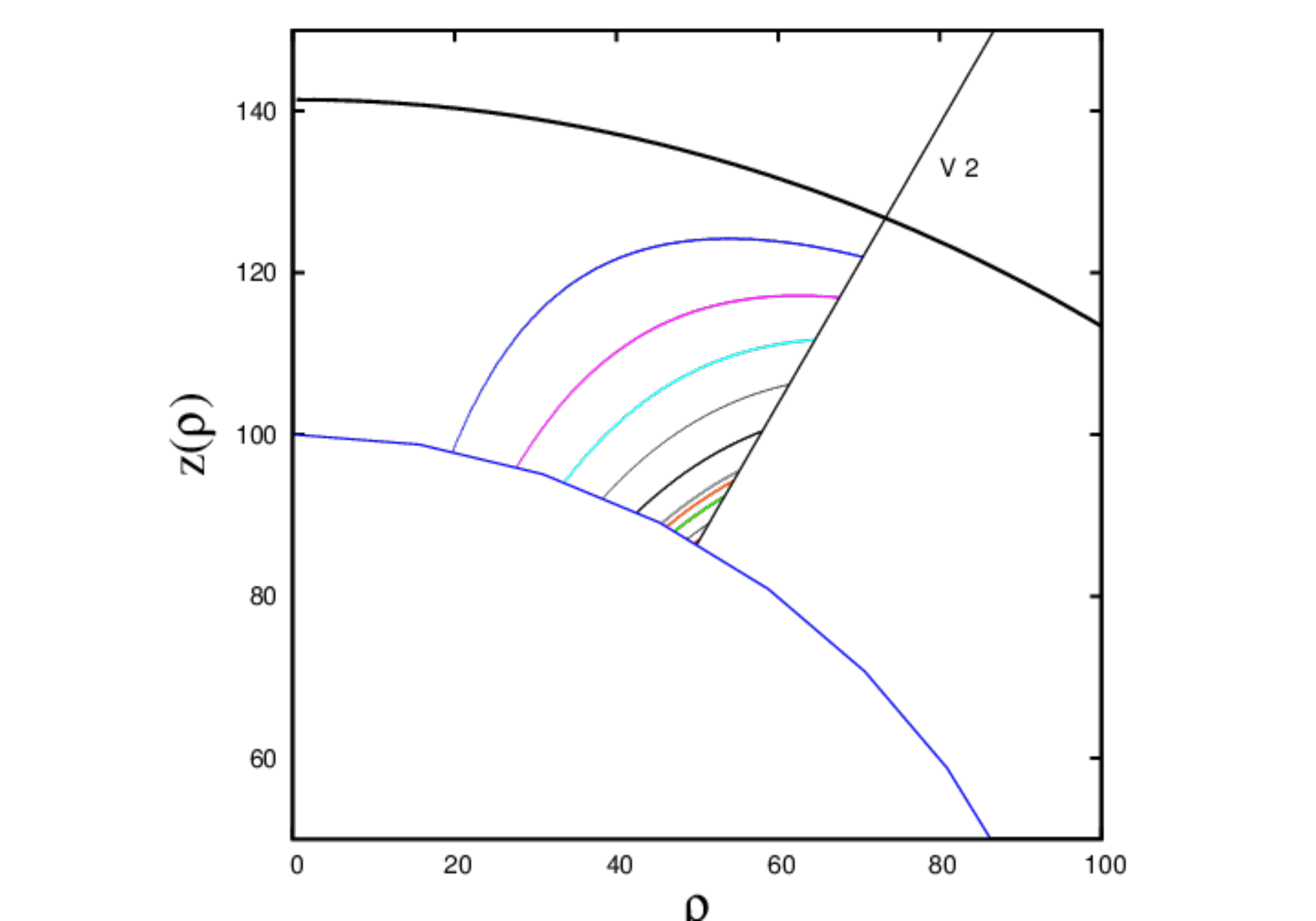}  
  \caption{\bf Some streamlines originating at different points at the
    termination shock are shown with different colors.  The Voyager-2
    trajectory is the black straight line, crossing the various
    streamlines on which the solar wind has evolved differently \citep{Fahr-etal-2016}.}
  \label{fig:1}
\end{figure}

{\bf Because the heliosheath plasma is convected along streamlines
  which originate at different points on the termination shock, it
  must be kept in mind, that Voyager-2 crosses different streamlines
  (see Fig.~\ref{fig:1}), when moving through the inner heliosheath.
  On these different streamlines the solar wind evolves differently,
  as has been quantified in \citet{Fahr-etal-2016}.}
Furthermore, in that paper we assumed that the locally prevailing
temperature is a $\kappa$-temperature
$T_{\kappa }$ resulting from an incompressible plasma flow
\begin{equation}\label{eq:8}
T_{\kappa }(s)=\frac{m}{2K}\Theta ^{2}(s)\frac{\kappa (s)}{\kappa (s)-3/2}
\end{equation}
In the following, after discussing the variation of $\kappa$ and $\kappa$-temperature
along the Voyager-2 trajectory, we derive
how the Maxwellian temperatures can be translated in $\kappa$-temperatures.

\section{Heliosheath plasma along the Voyager-2 trajectory}

\begin{figure}[t!]
  \includegraphics[width=0.9\columnwidth]{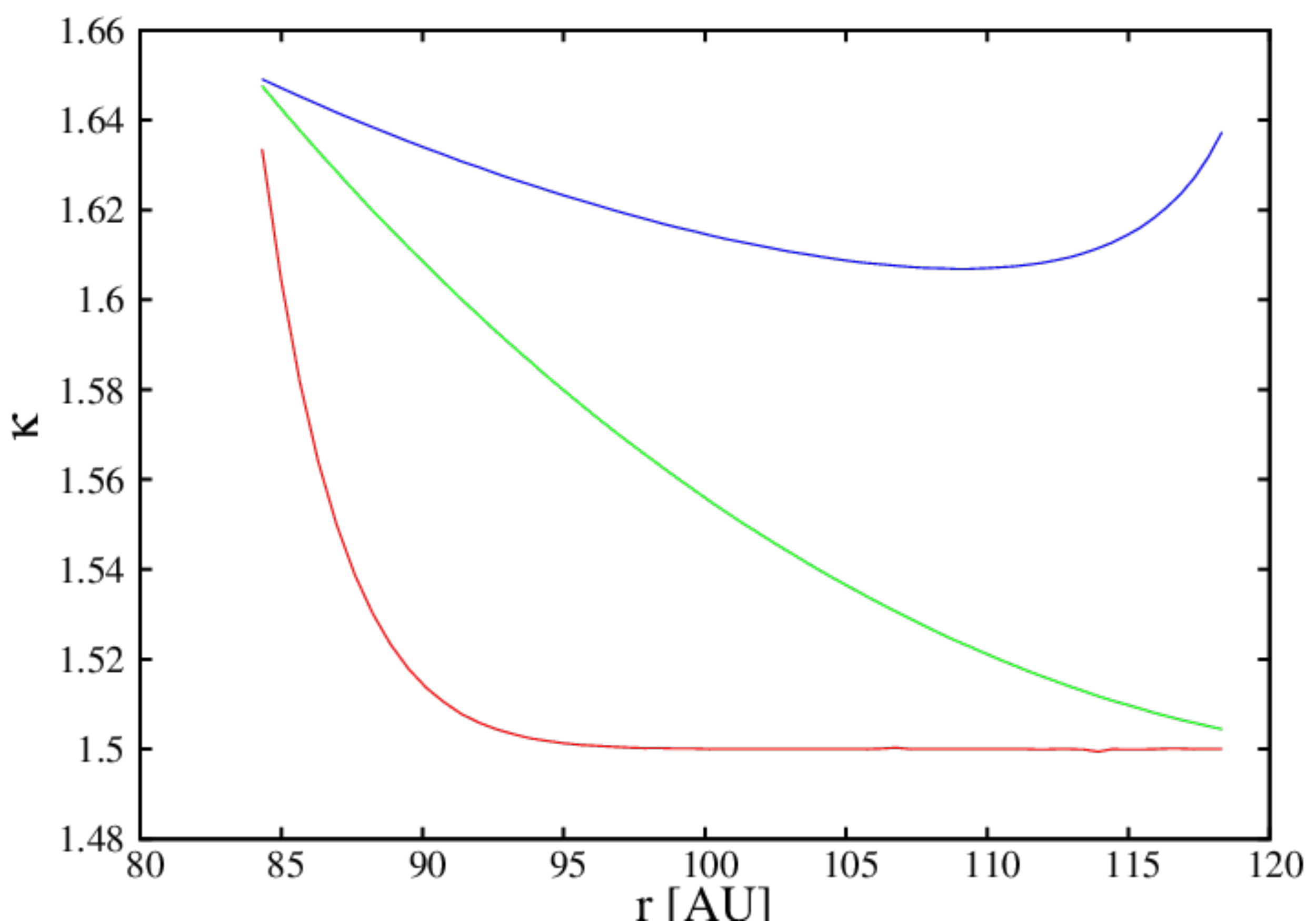}  
  \caption{$\kappa$-evolution for three different momentum diffusion
    coefficients: $D_0=10^{-10},10^{-9},10^{-8}$ (red green and blue
    curve respectively) along the Voyager-2 trajectory. See
    \citet{Fahr-etal-2016} for details.}
  \label{fig:2}
\end{figure}

\begin{figure}[t!]
  \includegraphics[width=0.9\columnwidth]{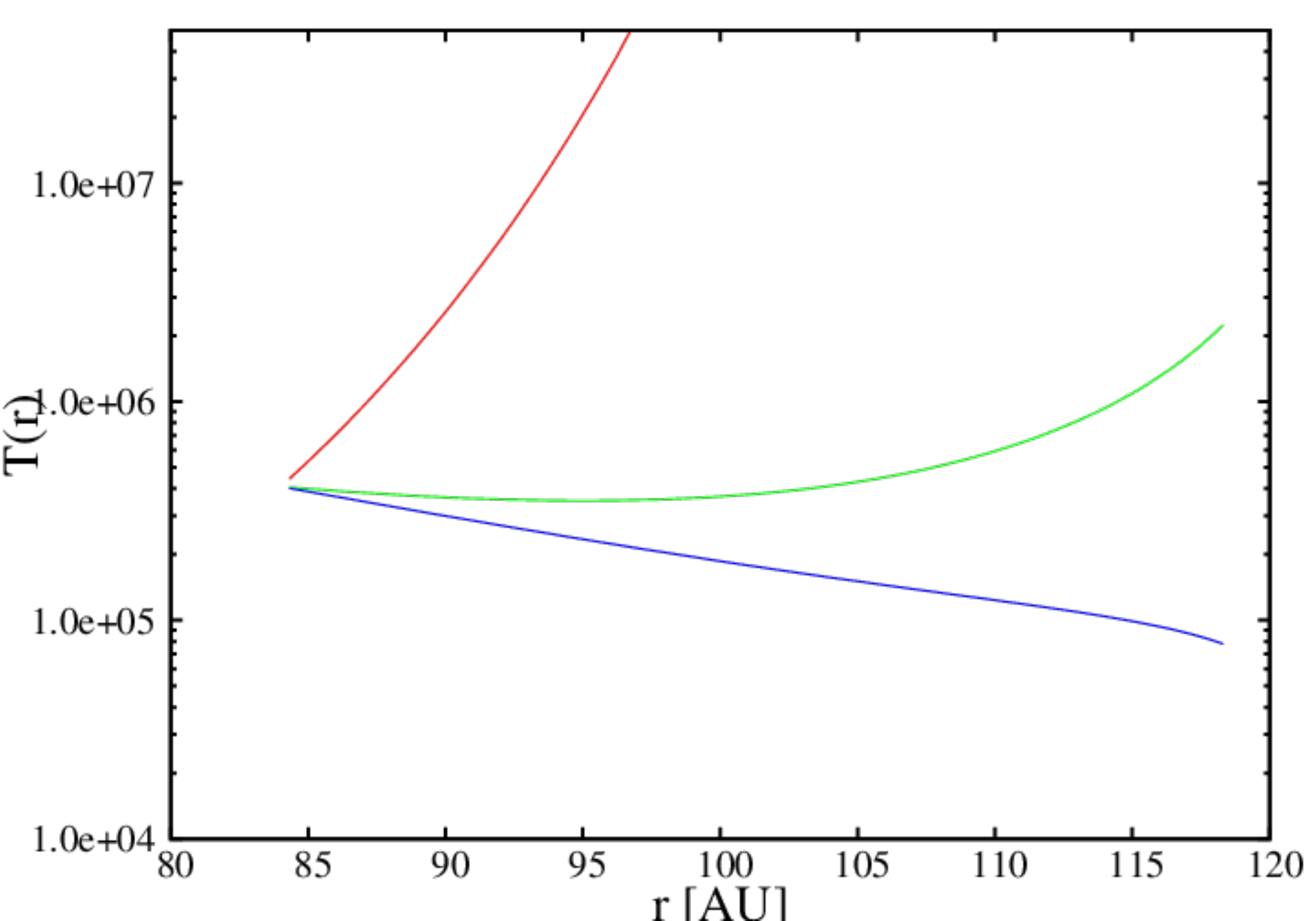}  
  \caption{The $\kappa$-temperature derived from Eq.~(9) of \citet{Fahr-etal-2016}. using
  the results shown in Fig~\ref{fig:2}}
  \label{fig:3}
\end{figure}

In \citet{Fahr-etal-2016} we have developed a numerical procedure to
calculate the ion pressure evolution along arbitrarily selected flow
lines of the heliosheath plasma flow on the basis of a pressure
transport equation in which it was assumed that at each location in the
plasma flow the underlying distribution function can be approximated
by a $\kappa$-distribution $f(s,v)=f_{\kappa }(s,v)$. This method allowed us
to estimate the evolution of the $\kappa$-parameter along each streamline.

In Fig.~\ref{fig:2} we show, how the $\kappa$-parameter varies, in turn, along the Voyager-2
trajectory (displayed as heliocentric distance in AU). Depending
on the magnitude of the velocity diffusion coefficient $D_{0}$
\citep{Fahr-etal-2016}, one can see along the Voyager-2 trajectory
different $\kappa$-values (see green, red, or blue curves). The more
efficient the velocity diffusion process is, the lower are the
resulting $\kappa$-values at Voyager-2. Directly connected with these
$\kappa$-values are the associated $\kappa$-temperatures $T_{\kappa }$ along the Voyager-2
trajectory (see Eq.~(\ref{eq:8})) as shown in  Fig.~\ref{fig:3}.

\section{$\ensuremath{\kappa}$-temperatures along the Voyager-2 trajectory}
We use the standard coordinate system for distribution functions,
namely the space coordinates in the solar frame and the velocity
coordinates in the rest frame (bulk frame) of the fluid. We assume
isotropic distributions as defined in section~1.1 above in each case.

\begin{figure}[t!]
  \includegraphics[width=0.45\textwidth]{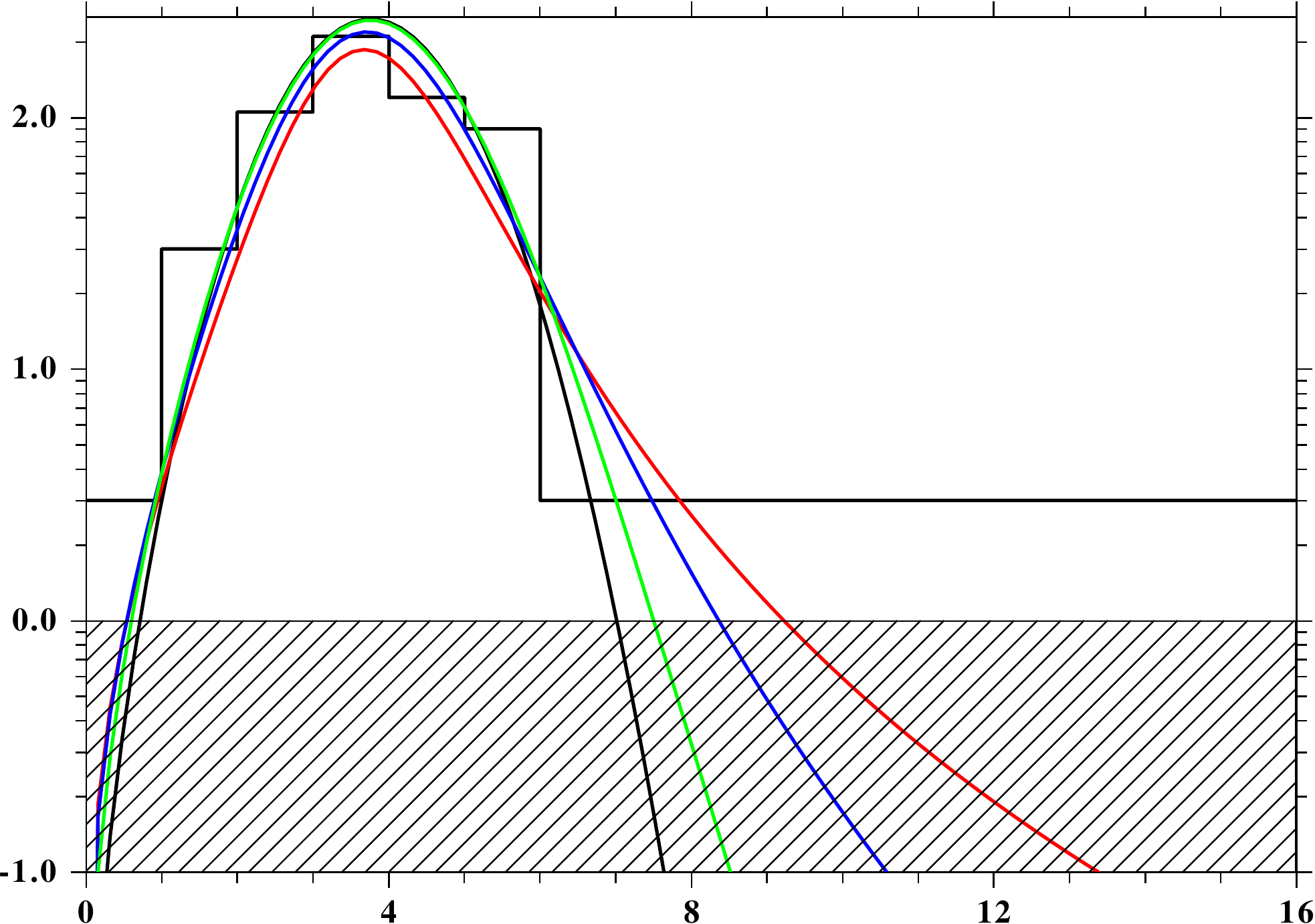}
  \caption{Sketch of a realistic situation for fitting the data
    (histogram, i.e.\ currents in the instrument cups). The
    black line shows a shifted Maxwellian and
    three shifted $\kappa$-distribution with $\kappa=1.6$ (red), $\kappa=3$ (blue), and
    $\kappa=10$ (green). The number on the x-axis are the channel
    numbers which can be translated to energy or velocity. The y-axis
    is an arbitrary units.The distribution functions are multiplied
    with the velocity, which is approximately a representation of the
    currents, see \citet{Bridge-etal-1977}.\label{fig:sketch}}
\end{figure}

\subsection{The data fitting procedure}
A problem now arises unavoidably when using observations, like those
from the plasma instrument on-board of Voyager-2 \citep{Richardson-2008}
where one has only a few data points in a limited energy range to fit
a distribution function .

The physical parameters, like density, temperature and velocity are
obtained from the currents of the different cups in the PLS-instrument
on-board of Voyager-2. These currents correspond to a kinetic energy
(momentum) from which the parameters are derived assuming a best fit
Maxwellian \citep{Richardson-2008,Barnett-1986}. Such best fit is
shown in Fig.~\ref{fig:sketch} using data given by
\citet{Richardson-2008}, in which the black (step-like) lines show the
measurements and the result of the best fit with a Maxwellian. We have
also plotted results using $\kappa$-distributions with different $\kappa$'s. It can be
seen that the representations by the alternative $\kappa$-distributions give similar
results in the measurement range but deviate from the Maxwellian fit
at higher energies (channels). So, in the range not covered by data
the tails of the distribution functions can contribute remarkably to
the higher moments, especially to the pressure (temperature). Thus the
latter can strongly differ from those obtained by using Maxwellians.

\subsection{Truncated distribution functions}

\begin{figure}[t!]
  \includegraphics[width=0.45\textwidth]{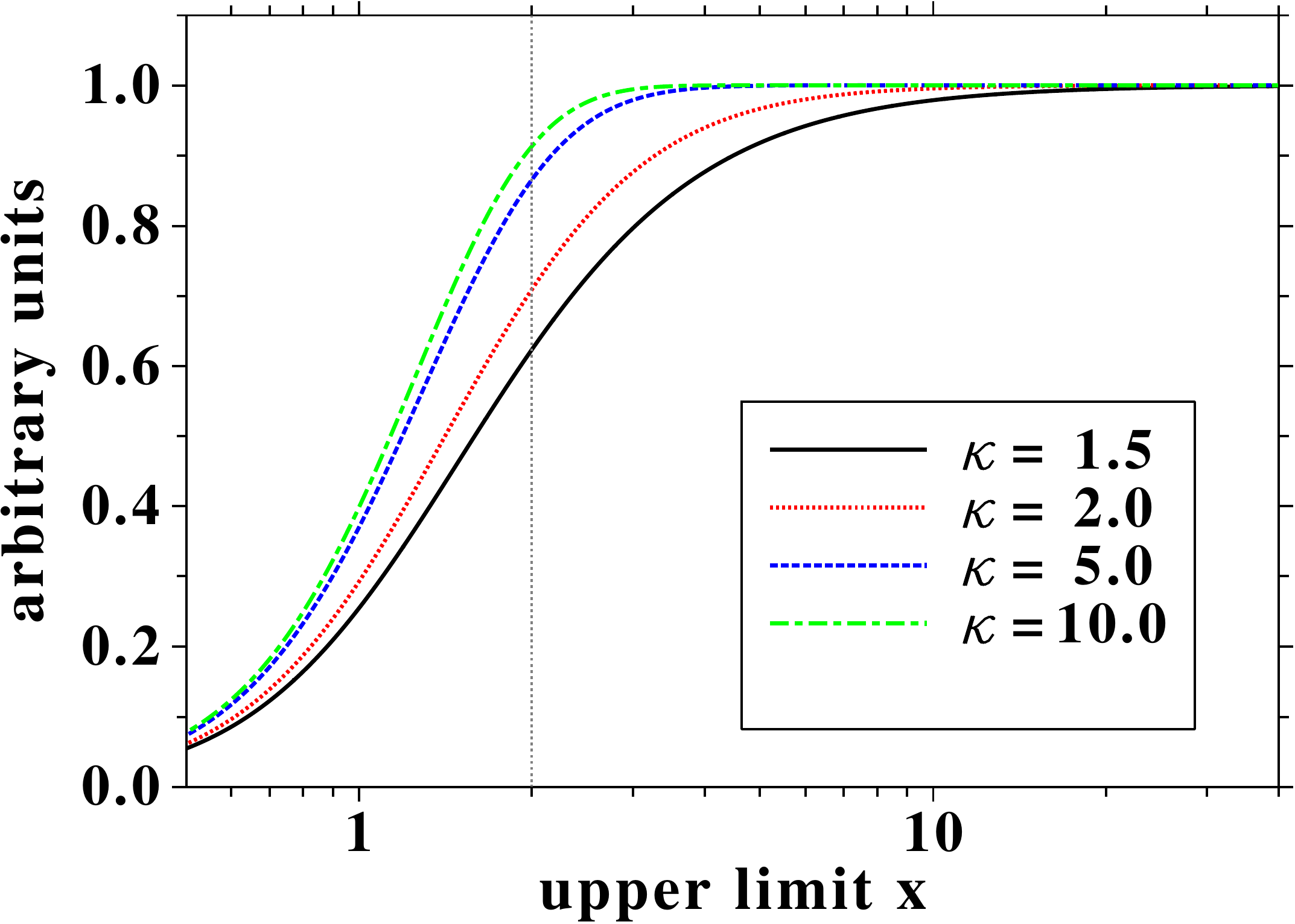}\hfill
  \includegraphics[width=0.45\textwidth]{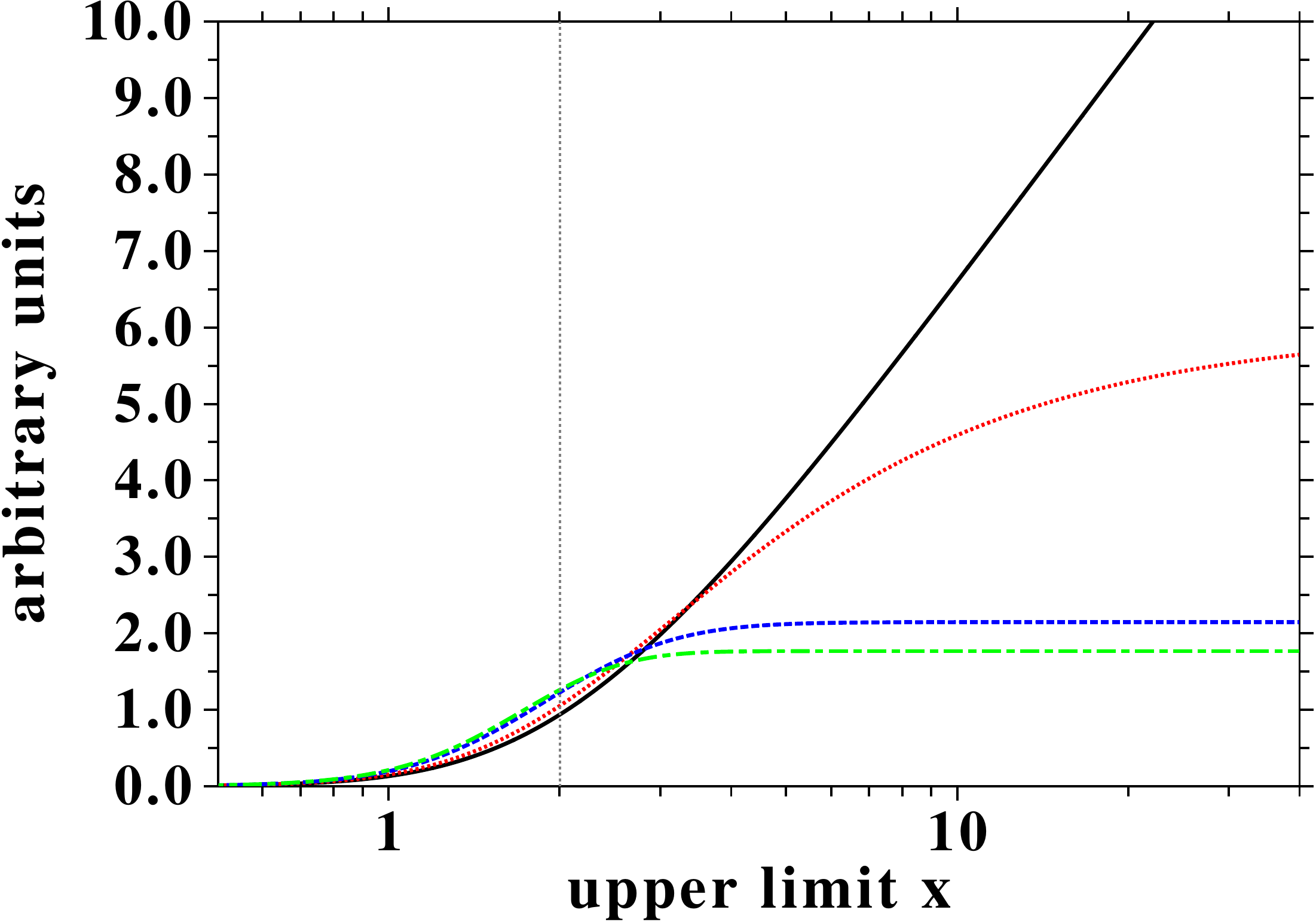}
  \caption{The  truncated moments $M^{0}_{\kappa,x}$ (upper panel) and
    $M^{2}_{\kappa,x}$ (lower panel) for different $\kappa$'s.  The
    logarithmic x-axis displays the speed normalized to 100\,km/s and
    the y-axis is in arbitrary units. See text for 
    details. The dotted vertical line indicates a speed of roughly
    200\,km/s, which is about the solar wind speed in the
    heliosheath. While a speed about 30 units corresponds to that of
    protons as measured in the lowest energy channel of the LECP
    instrument on-board of Voyager. 
    \label{fig:trunc}}
\end{figure}

Based on the above discussion we have only available truncated
distribution functions which have to deliver the fit. In the case of a
Maxwellian this is not a problem, but for the $\kappa$-distribution, because the core
for different $\kappa$ values can be fitted quite well, while the very
different tails, not taken into account into the fit, imply, that
particles far off the core contribute remarkably to the temperature
moments. It can be seen in Fig.~\ref{fig:sketch} that the fits to the
$\kappa$-distribution can easily produce substantial differences in the expected values
for different $\kappa$. This is shown in Fig.~\ref{fig:trunc} where we
have plotted the truncated zeroth and second order moments normalized
to the number density. That is:
 \begin{eqnarray}
   \label{eq:trucmom}
   M^{0}_{\kappa,x} = 4\pi\int\limits_{0}^{x}v^{2} f_{\kappa}dv \qquad
   M^{2}_{\kappa,x} = 4\pi\int\limits_{0}^{x}v^{4} f_{\kappa}dv
 \end{eqnarray}
 It can be seen in the upper panel of Fig.\ref{fig:trunc} that for
 different sets of ($\kappa,\Theta$)
 for small normalized speed values the different $\kappa$-distribution zeroth order
 moments are approximately the same, but strongly differ in the second
 order moments for small $\kappa \lessapprox 2$
 above a truncation speed (see lower panel), which is usually higher
 than the range covered by observations.  These incomplete
 distribution functions are the true problem when interpreting
 observational data, while the problem faced by
 \citet{Nicolaou-Livadiotis-2016} considering complete (idealized)
 distribution functions appears more academic. Measured distributions
 are constrained by the instrument limitations, for instance, given by
 the energy range detected by the instrument, where the highest energy
 channel may  define a truncation limit of the distribution
 function and its moments.
 
\begin{figure}[t!]
  \includegraphics[width=0.95\columnwidth]{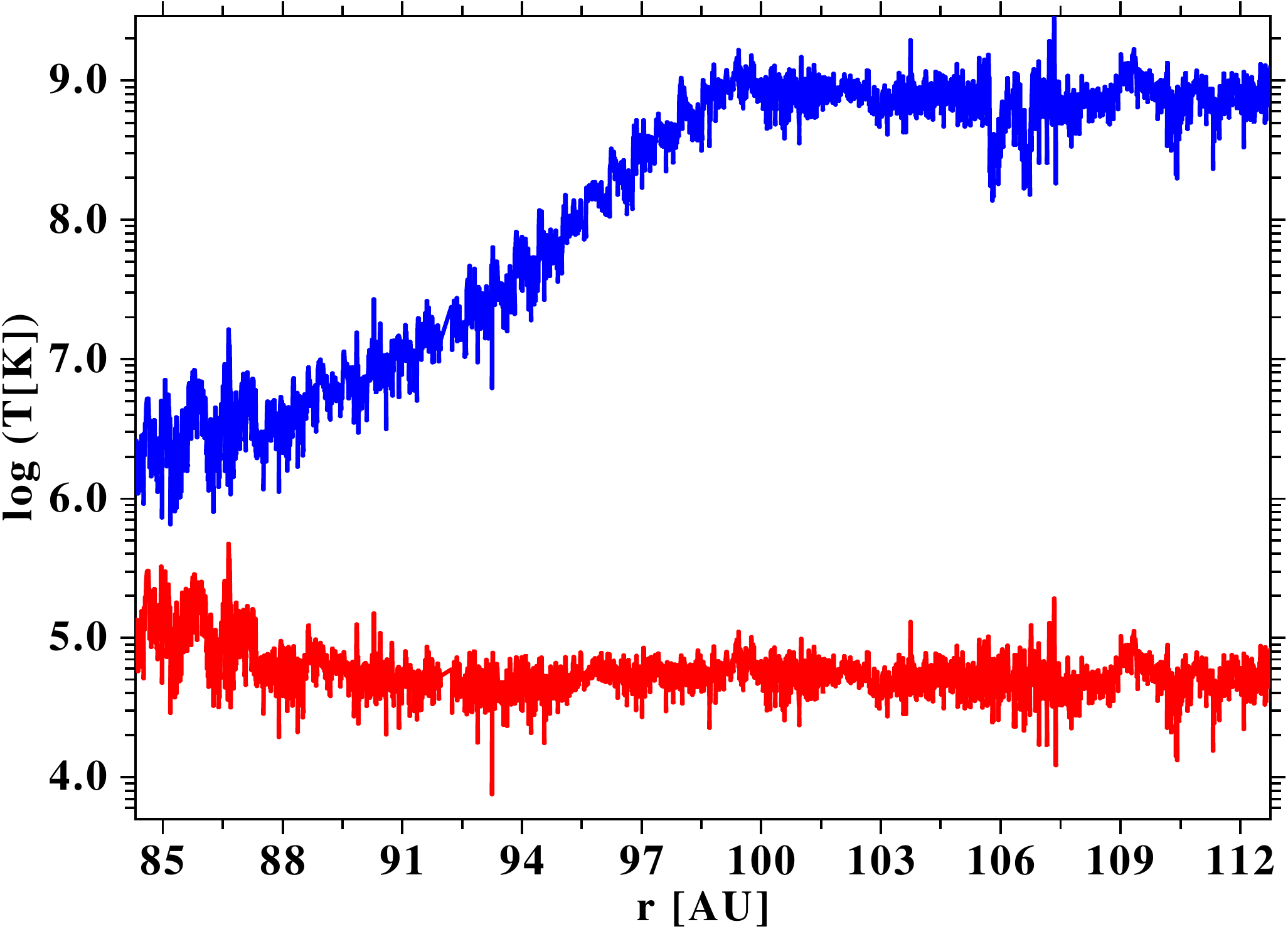}
  \includegraphics[width=0.95\columnwidth]{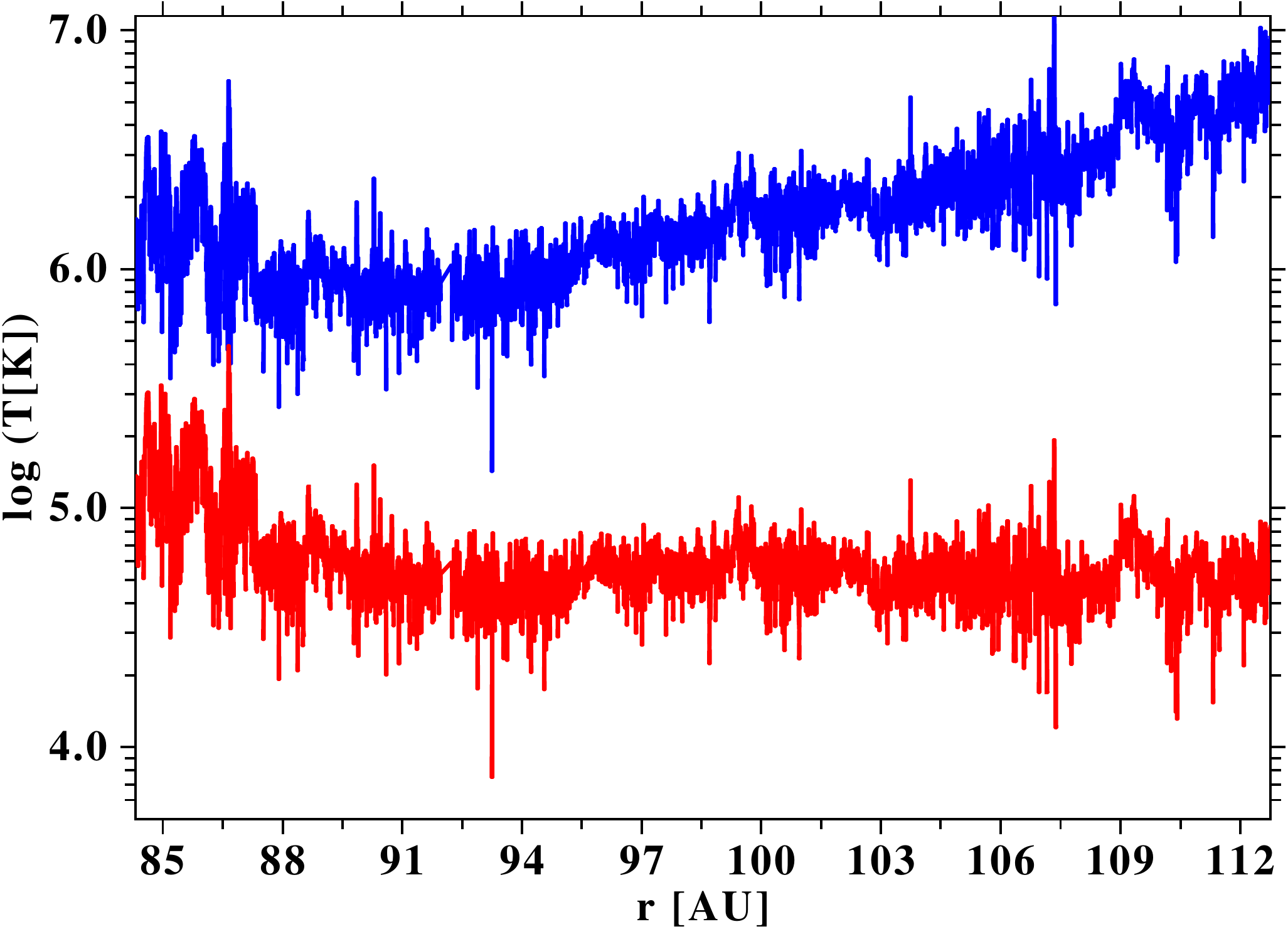}
  \includegraphics[width=0.95\columnwidth]{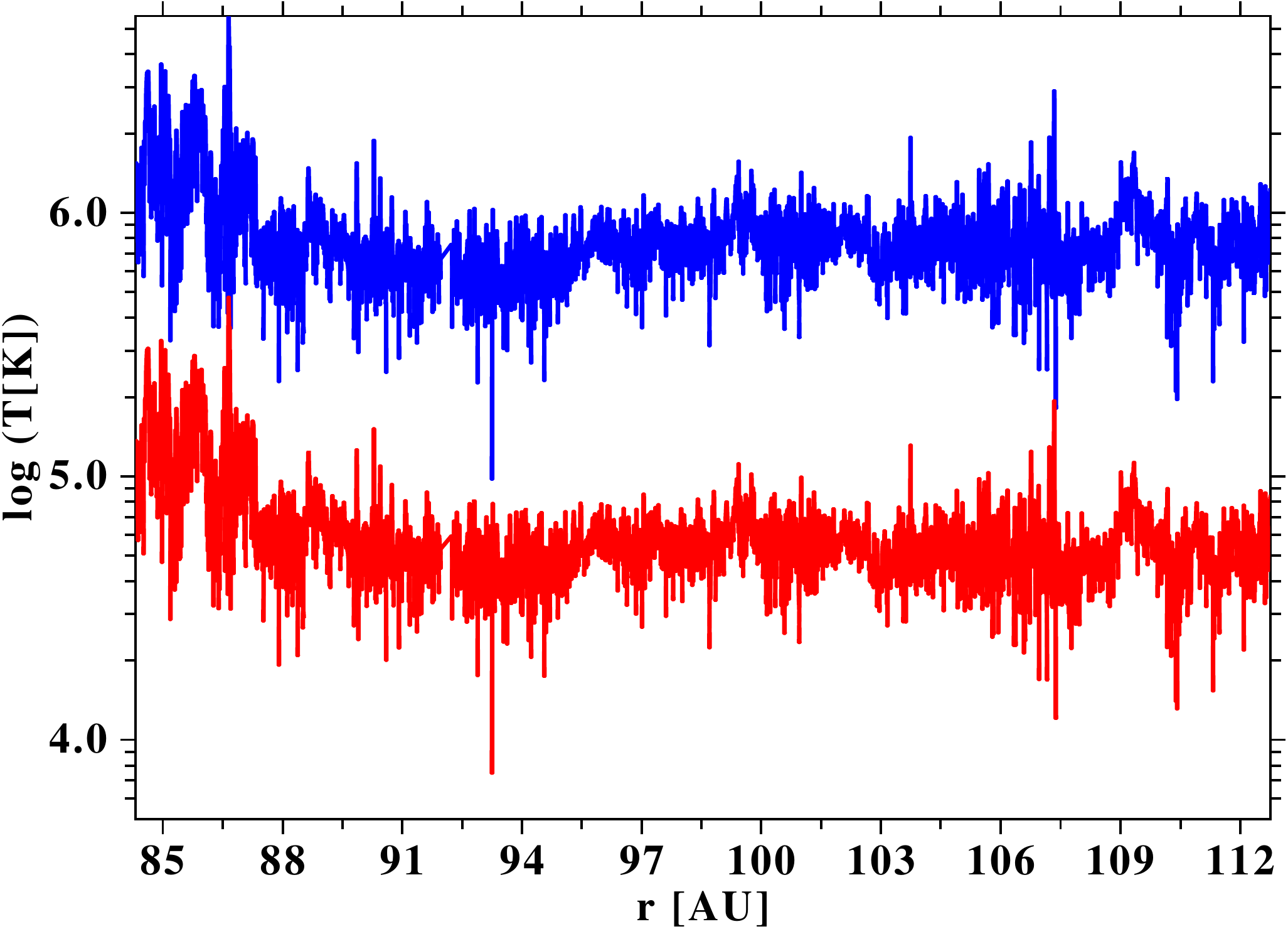}
  \caption{The estimated $\kappa$-temperature using Eq.~(\ref{eq:T-fin}) and the
    $\kappa$-parameter along the Voyager-2 trajectory as calculated by
    \citet{Fahr-etal-2016}. The lower red curves are the Voyager data
    form NSSDC
    (\url{http://omniweb.gsfc.nasa.gov/coho/form/voyager2.html}) From
    top to bottom, the diffusion coefficients as given in
    \citet{Fahr-etal-2016} change from
    $D_{0}=10^{-8},10^{-9}, 10^{-10}$\,cm$^{2}/s$. 
    \label{fig:tdata}}
\end{figure}

We have extended the range of the x-axis to match the lowest
channel of the LECP instrument on-board of Voyager-2, which measures
energetic protons in the energy range of $\approx$40\,keV
corresponding to roughly 3000\,km/s. It becomes obvious form
Fig.~\ref{fig:trunc} that these particles do not essentially
contribute to the number density for different $\kappa$'s, but they
heavily influence the second order moment (the pressure). Because the
contribution of the high energy particles to the number density is so
small, it is not an easy task to determine the $\kappa$-value. It will
require more advanced models, which go far beyond the scope of this
paper to determine the $\kappa$-distribution from data of different instruments.  

\subsection{Analytic estimate of the $\ensuremath{\kappa}$-distribution  temperature}

From the second moments we can find a relation between the pressures
of a $\kappa$-distribution $P_{\kappa}$ and that of a Maxwellian $P_{m}$ (see Eq.~(\ref{eq:p-ratio}))
\begin{eqnarray}
P_{\kappa} = \frac{2\kappa}{2\kappa -3} \, \frac{
  \Theta^{2}}{v_{p}^{2}} P_{m} 
\end{eqnarray}
Now we can either calculate the Maxwellian pressure from the
  observed Maxwellian temperature $T_{m}$ by
  \begin{eqnarray}
    P_{m} = n k_{B} T_{m}
  \end{eqnarray}
  with the Boltzmann constant $k_{B}$ and comparing the pressure terms
  or  we define the
  $\kappa$-temperature $T_{\kappa}$ by:
  \begin{eqnarray}
    T_{\kappa} = \frac{P_{\kappa}}{n k_{B}}
  \end{eqnarray}
 and compare the ``temperatures''. The result is the same, except for
  absolute numbers. With this definition we find:
\begin{eqnarray}
  \label{eq:T-ratio}
T_{\kappa} = \frac{2\kappa}{2\kappa -3} \, \frac{
  \Theta^{2}}{v_{p}^{2}} T_{m}   
\end{eqnarray}
which is identical to Eq.~(\ref{eq:8}).

Furthermore, because for $\kappa\rightarrow\infty$ the
$\kappa$-temperature must be equal to that of the  Maxwellian,
we find that $v_{p}=\Theta$ and finally, since $\Theta$ is independent
of $\kappa$,
\begin{eqnarray}
\label{eq:T-fin}
T_{\kappa} = \frac{2\kappa}{2\kappa -3} T_{m}   
\end{eqnarray}
depends only on $\kappa$ and can be easily calculated from the
Maxwellian temperature knowing the corresponding $\kappa$.  Thus one
can use the temperatures derived from a Maxwellian distribution and
calculate from that the $\kappa$-temperatures, where $\kappa$ and $\Theta$ can be
estimated elsewhere, for instance by solving a kinetic transport
equation as done in \citet{Fahr-etal-2016}.

That means also that the presented temperatures derived with a
Maxwellian as available from the Omniweb
(\url{http://omniweb.gsfc.nasa.gov/coho/form/voyager2.html}) must be
taken with care, because they should translate into $\kappa$-distribution temperatures
along the Voyager-2 trajectory through the inner heliosheath.  With
formula~(\ref{eq:T-fin}) this is an easy task knowing $\kappa$.

In Fig.~\ref{fig:tdata} we have plotted the Voyager-2 temperature data
between 84 to 112\,AU, together with those obtained from the $\kappa$-parameter of
\citet{Fahr-etal-2016} using Eq.~(\ref{eq:T-fin}).  In the upper panel
of Fig.~\ref{fig:tdata} the $\kappa$-parameter reaches the limiting value of 1.5 and,
to avoid numerical problems it was set to $\kappa=1.5001$ beyond
$r \approx 98$\,AU. Thus beyond 98\,AU the model by
\citet{Fahr-etal-2016} is not applicable for such high values of the
diffusion coefficient ($D_{0}\approx10^{-8}$\,cm$^{2}$/s), while it can
nicely explain a non-adiabatic behavior of the temperature between
$\approx 88$ and $\approx 98$\,AU.

This shows, that the temperature highly depends on the underlying
distribution function, and thus temperature data have to be taken with
care, because they can easily lead to erroneous interpretations of the
data. This holds true in general.

\section{Conclusion}
The above study shows how Maxwellian temperatures derived from Voyager-2
measurements can easily be translated into corresponding $\kappa$-temperatures. We have
also shown \citep[like][]{Nicolaou-Livadiotis-2016} that the
temperature of a plasma highly depends on the underlying distribution
function and differs from that obtained by a Maxwellian.  This implies
the need of the knowledge of the underlying distribution function,
because, otherwise, the temperature is not well defined.

The recently presented study by \citet{Nicolaou-Livadiotis-2016} is
not helpful in the ``data-relevant'' aspects discussed here, because
the authors compare moments calculated on the basis of $\kappa$- or
Maxwell-distributions, however, taken from an infinite velocity range,
while in reality data only support moments in a very limited velocity
range. This is the important aspect that has to be taken serious in
these matters.

Nevertheless, after deriving a $\kappa$-distribution from theoretical considerations it
is an easy task to determine the corresponding $\kappa$-temperature when a Maxwellian
temperature is given.  From IBEX observations
\citep{McComas-etal-2015} it might be possible to obtain the
$\kappa$-value
from which the temperature can be estimated.
 Note, that the previous attempts to do so
  \citep[e.g.\ ][]{Livadiotis-etal-2011,Zirnstein-McComas-2015} use a
  $\kappa$-independent
  temperature, which is a concept under debate
  \citep{Lazar-etal-2016, Lazar-etal-2017}.

The procedure discussed in the present paper, may not only be applied
to spacecraft observations but to all observations in which the
temperature is derived with the assumption that the underlying
distribution function is a Maxwellian.

~\\

\appendix
\section{Moments of the distribution function}
From the moments of a distribution function the macroscopically 
observable quantities can be derived
\citep[for example][and many others]{Goedbloed-Poedts-2004}. To
calculate the moments one has to integrate the distribution function
times some power $\alpha$ of the speed. For the Maxwell distribution the integrals
are well known and can be found elsewhere
\citep[e.g.\ ][]{Gradshteyn-Ryzhik-2007}. For the
$\kappa$-distribution they can also be found in a more general form in 
\citet{Gradshteyn-Ryzhik-2007}, Nr 3.251.2:
\begin{eqnarray}
  \label{eq:5}
  I &=& \int\limits_{0}^{\infty} v^{\mu-1}(1+\beta x^{p})^{-\nu}
  \\\nonumber
    &=& 
 \frac{1}{p}\beta^{-\mu/p}
        \frac{\Gamma(\mu/p)\Gamma(\nu-\mu/p)}{\Gamma(\nu)}\\\nonumber
\end{eqnarray}
where $\Gamma$ is the Gamma function. 
With
\begin{eqnarray}
  \mu = 3 + \alpha, \qquad \beta = \frac{1}{\kappa \Theta^{2}}, \qquad
  \nu = \kappa + 1, \qquad p = 2  
\end{eqnarray}
we have the integral \citep{Fahr-etal-2014}
\begin{eqnarray}
  \label{eq:6}
  I_{\alpha} &\equiv& \int\limits_{0}^{\infty}v^{\alpha+2}\left(1+\frac{ v^{2}}{\kappa\Theta^{2}}\right)^{-\kappa-1}dv \\\nonumber
   &=& \frac{1}{2}\left(\sqrt{\kappa}
  \Theta^{2}\right)^{\frac{3+\alpha}{2}}
  \frac{\Gamma\left(\frac{3+\alpha}{2}\right) \Gamma\left(\kappa-\frac{1+\alpha}{2}\right)}{\Gamma(\kappa+1)}
\end{eqnarray}
and thus $I_{0}\ g(\kappa,\Theta)=1$, {\bf where $g(\kappa,\Theta)=n(\vec{r}) \dfrac{\Gamma(\kappa+1)}{
  \left(\sqrt{\pi}\sqrt{\kappa}\Theta\right)^{3}\Gamma\left(\kappa-\dfrac{1}{2}\right)}$
is the normalisation factor of the $\kappa$-distribution defined in Eq.~\ref{eq:3}.}

With the above integral we thus can easily calculate the following
moments of the $\kappa$-distribution 
  \begin{align}\nonumber
    M^{0}&= \int f_{\kappa}(\vec{r},\vec{v},t)
    d^{3}v &\quad\mathrm{number\ density}\\\nonumber
    \vec{M}^{1}&= \dfrac{1}{n(\vec{r},t)}\int\vec{v} f_{\kappa}(\vec{r},\vec{v},t)
    d^{3}v = \vec{0}&\quad \mathrm{velocity}\\\nonumber
    \overset{\leftrightarrow}{M}^{2}&= \int\vec{v}\otimes\vec{v} f_{\kappa}(\vec{r},\vec{v},t)
    d^{3}v &\quad \mathrm{stress\ tensor\ per\ unit\ mass}
  \end{align}
  of $f_{\kappa}$ and compare them to those for a Maxwell
  distribution $f_{m}$. In the comoving reference frame the first moment
  $\vec{M}^{1}$ vanishes. Furthermore, we assume that the stress
  tensor will be described by an isotropic pressure and thus the dyadic
  $\vec{v}\otimes\vec{v}$ can be contracted to a scalar $M^{2}$ as
$\vec{v}\otimes\vec{v}\rightarrow v^{2}$. In addition we calculate the
most probable speed by 
\begin{eqnarray}
  M^{1} = \dfrac{1}{n(\vec{r},t)}\int v f(\vec{r},\vec{v},t) d^{3}v = v_{p}
\end{eqnarray}

We find for the different moments:
  \begin{align}
    M^{0}_{m}&=M^{0}_{\kappa} = n(\vec{r},t)\\
    M^{1}_{m}&=\frac{2}{\sqrt{\pi}}\sqrt{\frac{2k T}{m}} 
\equiv\frac{2}{\sqrt{\pi}} v_{p}\\
 M_{m}^{2} &=\frac{3}{2} v_{p}^{2}n(\vec{r},t)\\
  M_{\kappa}^{1} &= \frac{2}{\sqrt{\pi}} \Theta \frac{1}{(\kappa-1)\sqrt{\kappa}}\frac{\Gamma\left(\kappa+1\right)}{\Gamma\left(\kappa-\dfrac{1}{2}\right)} \\
  M_{\kappa}^{2} &=\frac{3\kappa \Theta^{2}}{2\kappa -3}n(\vec{r},t)
\end{align}

The pressure ratio is:
\begin{eqnarray}
  \label{eq:p-ratio}
P_{\kappa} = \frac{2\kappa}{2\kappa -3} \, \frac{
  \Theta^{2}}{v_{p}^{2}} P_{m} 
\end{eqnarray}

Finally, replacing the velocity in the fluid rest frame by that in the
observer frame, that is: $\vec{v}=\vec{u}-\vec{w}$, where $\vec{u}$ is
the fluid bulk velocity and $\vec{w}$ the thermal velocity, one easily
finds that the zeroth moment remains, the first moment gives the bulk
speed and in the second moment the ram pressure appears as an additional
term.

\begin{acknowledgements}
  The work of HF and KS was partly carried out within the framework of
  the bilateral BMBF-NRF-project ``Astrohel'' (01DG15009) funded by
  the Bundesministerium f\"ur Bildung und Forschung. The
  responsibility of the contents of this work is with the authors. AS
  acknowledges the support via the DFG project FI706/21-1. JDR was
  supported under NASA contract 959203 from the Jet Propulsion
  Laboratory to the Massachusetts Institute of Technology.
\end{acknowledgements}


\end{document}